\documentclass[twocolumn,showpacs,preprintnumbers,nofootinbib,prl,superscriptaddress,groupedaddress,10pt,aps]{revtex4-1}
\usepackage{graphicx,amssymb,amsmath,amsthm,amsfonts,epsfig,epsf}
\usepackage[linktocpage]{hyperref}
\usepackage[usenames]{color}
\usepackage{epstopdf}

\usepackage{bm}	
\usepackage{dcolumn}
\usepackage[latin1]{inputenc}
\usepackage{latexsym}
\usepackage{rotating}
\usepackage{hyperref}
\usepackage{color}
\usepackage{longtable}
\usepackage{enumerate}
\usepackage{tensor}
\usepackage{url}
\def\nn{\nonumber}

\newcommand{\ben}{\begin{enumerate}}
\newcommand{\een}{\end{enumerate}}

\def\be{\begin{equation}}
\def\ee{\end{equation}}
\def\bea{\begin{eqnarray}}
\def\eea{\end{eqnarray}}
\def\nn{\nonumber}
\newcommand{\beq}{\begin{eqnarray}}
\newcommand{\eeq}{\end{eqnarray}} 
\newcommand{\ba}{\begin{align}}
\newcommand{\ea}{\end{align}}

\begin{document}

% \newpage 

% \includepdf{erratum_arXiv.eps}
% \includepdf[page=1]{erratum_arXiv.pdf}
\epsfig{file=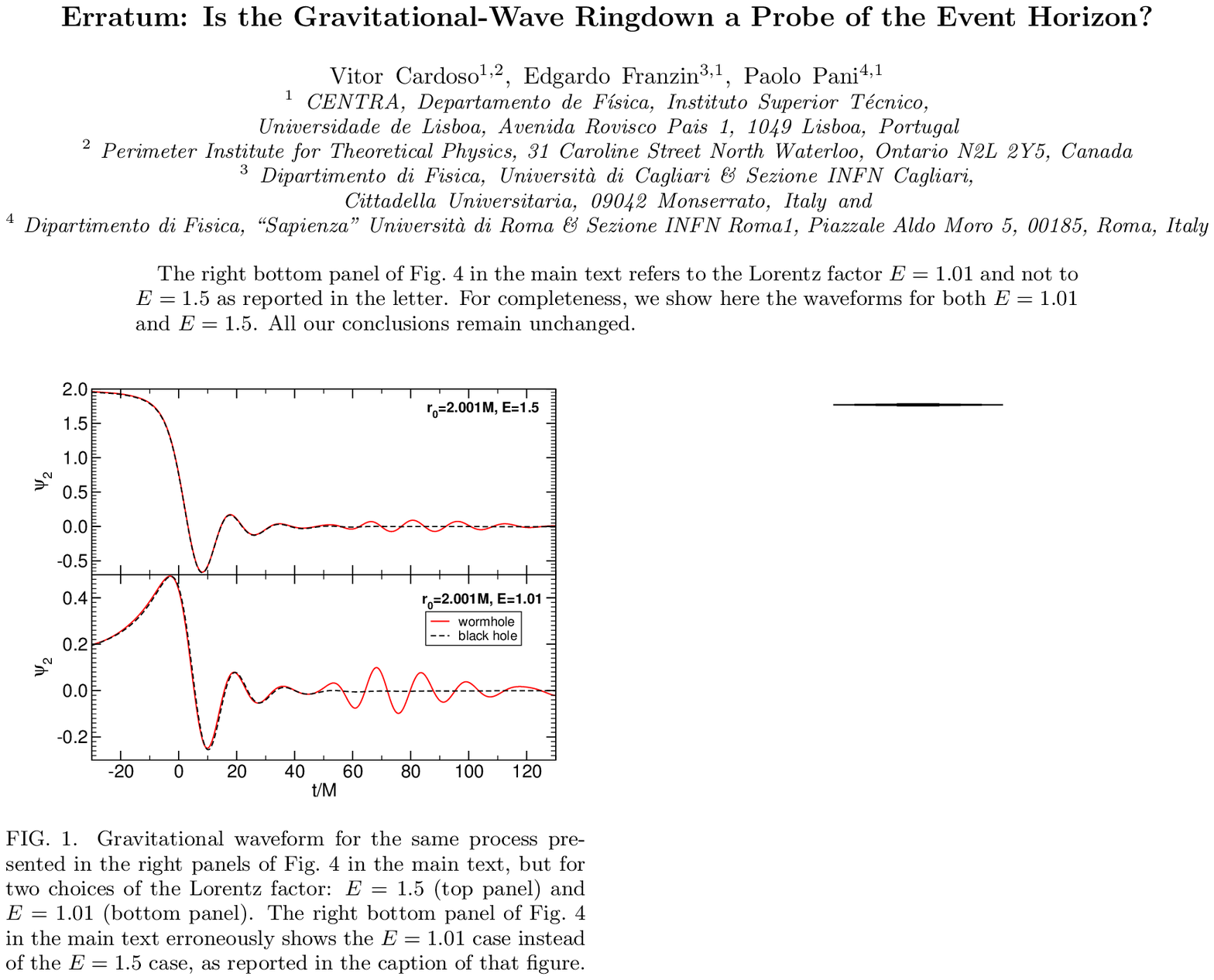,width=1.1\textwidth,angle=0,clip=true}

% \pagebreak
% \newpage 
% \clearpage

\clearpage

\title{Is the Gravitational-Wave Ringdown a Probe of the Event Horizon? }

\author{
Vitor Cardoso$^{1,2}$, % \footnote{Electronic address: vitor.cardoso@tecnico.ulisboa.pt},
Edgardo Franzin$^{3,1}$,
Paolo Pani$^{4,1}$
}
\affiliation{${^1}$ CENTRA, Departamento de F\'{\i}sica, Instituto Superior T\'ecnico, Universidade de Lisboa, Avenida~Rovisco Pais 1, 1049 Lisboa, Portugal}
\affiliation{${^2}$ Perimeter Institute for Theoretical Physics, 31 Caroline Street North
Waterloo, Ontario N2L 2Y5, Canada}
\affiliation{${^3}$ Dipartimento di Fisica, Universit\`a di Cagliari \& Sezione INFN Cagliari, Cittadella Universitaria, 09042 Monserrato, Italy}
\affiliation{${^4}$ Dipartimento di Fisica, ``Sapienza'' Universit\`a di Roma \& Sezione INFN Roma1, Piazzale Aldo Moro 5, 00185, Roma, Italy}
%
% 
% \author{Vitor Cardoso}
% \affiliation{CENTRA, Departamento de F\'{\i}sica, Instituto Superior T\'ecnico, Universidade de Lisboa, Avenida~Rovisco Pais 1, 1049 Lisboa, Portugal}
% \affiliation{Perimeter Institute for Theoretical Physics, 31 Caroline Street North Waterloo, Ontario N2L 2Y5, Canada}
% %%%%
% \author{Edgardo Franzin}
% \affiliation{Dipartimento di Fisica, Universit\`a di Cagliari \& Sezione INFN Cagliari, Cittadella Universitaria, 09042 Monserrato, Italy}
% \affiliation{CENTRA, Departamento de F\'{\i}sica, Instituto Superior T\'ecnico, Universidade de Lisboa, Avenida~Rovisco
%   Pais 1, 1049 Lisboa, Portugal}
% %%%%%
% \author{Paolo Pani}
% \affiliation{Dipartimento di Fisica, Universit\`a di Roma ``La Sapienza'' \& Sezione INFN Roma1, P.A. Moro 5, 00185,
%   Roma, Italy}
% \affiliation{CENTRA, Departamento de F\'{\i}sica, Instituto Superior T\'ecnico, Universidade de Lisboa, Avenida~Rovisco Pais 1, 1049 Lisboa, Portugal}

\begin{abstract}
It is commonly believed that the ringdown signal from a binary coalescence provides a conclusive proof for the formation of an event horizon after the merger. This expectation is based on the assumption that the ringdown waveform at intermediate times is dominated by the quasinormal modes of the final object. We point out that this assumption should be taken with great care, and that very compact objects with a light ring will display a similar ringdown stage, even when their quasinormal-mode spectrum is completely different from that of a black hole. In other words, universal ringdown waveforms indicate the presence of light rings, rather than of horizons.
Only precision observations of the late-time ringdown signal, where the differences in the quasinormal-mode spectrum eventually show up, can be used to rule out exotic alternatives to black holes and to test quantum effects at the horizon scale.
\end{abstract}


\maketitle

%%%%%%%%%%%%%%%%%%%%%%%%%%%%%%%%%%%%%%%%%%%%%%%%%%%%%%%%%%%%%%%%%%%%%%%%%%%%%%
\noindent{\bf{\em I. Introduction.}}
%%%%%%%%%%%%%%%%%%%%%%%%%%%%%%%%%%%%%%%%%%%%%%%%%%%%%%%%%%%%%%%%%%%%%%%%%%%%%%
The first direct gravitational-wave (GW) detection of a compact-binary coalescence by aLIGO~\cite{Abbott:2016blz} opens up the exciting possibility of testing gravity in extreme regimes~\cite{TheLIGOScientific:2016src,Yunes:2013dva,Berti:2015itd}.
The detected GW signal is characterized by three phases~\cite{Buonanno:2006ui,Berti:2007fi,Sperhake:2011xk}: the inspiral stage, corresponding to large separations and well approximated by post-Newtonian theory; the merger phase when the two objects coalesce and which can only be described accurately through numerical simulations; and the ringdown phase when the
merger end-product relaxes to a stationary, equilibrium solution of the field equations~\cite{Sperhake:2011xk,Berti:2009kk,Blanchet:2013haa}.

It is commonly believed\footnote{As far as we are aware, Refs.~\cite{Damour:2007ap,Barausse:2014tra} discuss this issue correctly for the first time (cf. also a related discussion in Ref.~\cite{Nakamura:2016gri}).} that the ringdown waveform is dominated by the quasinormal modes (QNMs)~\cite{Kokkotas:1999bd,Berti:2009kk,Konoplya:2011qq} of the final object. If the latter is a Kerr black hole (BH), the entire QNM spectrum is characterized only by the BH mass and angular momentum. Thus, the detection of a few modes from the ringdown signal can allow for precision measurements of the BH mass and spin, and possibly of higher multipole moments, which can be used to perform null-hypothesis tests of the no-hair theorems of general relativity~\cite{Berti:2005ys,Berti:2007zu,Gossan:2011ha,Berti:2015itd}. This reasoning suggests that the GW ringdown signal provides a way to prove the existence of an event horizon in dark, compact objects. In light of the intrinsic limitations that inevitably plague any electromagnetic test of an event horizon, ringdown detections might arguably provide the \emph{only} conclusive proof of the existence of BHs~\cite{Abramowicz:2002vt}.

%%%%%%%%%%%%%%%%%%%%%%%%%%%%%%%%%%%%%%%%%%%%%%%%%%%%%%%%%%%%%%%%%%%%%%%%%%%%%%
\noindent{\bf{\em II. Light ring, ringdown and QNMs.}}
%%%%%%%%%%%%%%%%%%%%%%%%%%%%%%%%%%%%%%%%%%%%%%%%%%%%%%%%%%%%%%%%%%%%%%%%%%%%%%
The argument above relies on the assumption that the ringdown modes coincide with the QNM frequencies, defined as the poles of the appropriate Green's function in the complex plane~\cite{Berti:2009kk}. We stress that this correspondence does not hold in general. The QNMs of a BH are intimately related to the peculiar boundary conditions required at the event horizon, namely absence of outgoing waves. 
If the final object does not possess a horizon, the boundary conditions change completely, thus drastically affecting the QNM structure. 
On the other hand, the ringdown waves of the distorted compact object are closely related to the null, unstable, geodesics in the 
spacetime~\citep{Press:1971wr,1972ApJ...172L..95G,Ferrari:1984zz,Cardoso:2008bp,Berti:2009kk}, their frequency and damping time being associated with the orbital frequency and with the instability time scale of circular null geodesics, respectively. Thus, in principle, the ringdown phase should not depend on the presence of a horizon as long as the final object has a light ring. 

If the final object is a BH, the ingoing condition at the horizon simply takes the ringdown waves and ``carries'' them inside the BH. In this case, the BH QNMs \emph{incidentally} describe also the ringdown phase.
However, if the horizon is replaced by a surface of different nature (as, e.g., in the gravastar~\cite{Mazur:2004fk} or in the firewall~\cite{Almheiri:2012rt} proposals) the relaxation of the corresponding horizonless compact object should then consist on the usual light-ring ringdown modes (which are no longer QNMs), followed by the proper modes of vibration of the object itself. The former are insensitive to the boundary conditions and similar to the BH case, whereas the latter (which one usually refers to as QNMs) can differ dramatically from their BH counterpart, since they are defined by different boundary conditions.

%%%%%%%%%%%%%%%%%%%%%%%%%%%%%%%%%%%%%%%%%%%%%%%%%%%%%%%%%%%%%%%%%%%%%%%%%%%%%%
\noindent{\bf{\em III. Setup.}}
%%%%%%%%%%%%%%%%%%%%%%%%%%%%%%%%%%%%%%%%%%%%%%%%%%%%%%%%%%%%%%%%%%%%%%%%%%%%%%
%
\begin{figure}[th]
\begin{center}
\epsfig{file=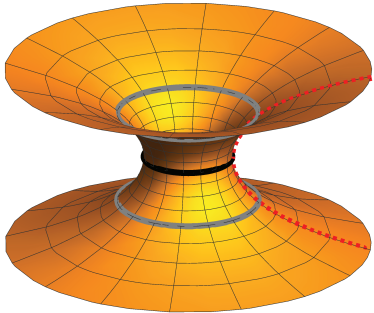,width=0.3\textwidth,angle=0,clip=true}
\caption{
Illustration of a dynamical process involving a compact horizonless object. A point particle plunges radially (red dashed curve) in a wormhole spacetime, and emerges in another ``universe''. 
The black curve denotes the wormhole's throat, the two gray curves are the light rings. When the particle crosses each of these curves, it excites characteristic modes which are trapped between the light-ring potential wells, see Figs.~\ref{fig:potential} and~\ref{fig:ringdown}.
\label{fig:embedding}}
\end{center}
\end{figure}
To the best of our knowledge, the above picture has never been verified in the context of GW tests of an event horizon. Here we perform such analysis by considering the ringdown signal and the QNMs associated with a horizonless compact object with a light ring. 
For definiteness\footnote{The main qualitative features of our analysis are independent of the specific horizonless object and apply also to spherical shells of matter, gravastars, compact boson stars and others~\cite{Leung:1999rh,Barausse:2014tra,Barausse:2014pra,Cardoso:2014sna}.}, we focus on the gravitational radiation emitted by a point particle in radial motion towards a traversable wormhole~\cite{Morris:1988tu,VisserBook} (cf. Fig.~\ref{fig:embedding} for an illustration). 

The specific solution is obtained by identifying two Schwarzschild metrics with the same mass $M$ at the throat $r=r_0>2M$ (we use $G=c=1$ units). In Schwarzschild coordinates, the two metrics are identical and described by $ds^2=-F dt^2+F^{-1} dr^2+r^2d\Omega^2$, where $F=1-2M/r$. Because Schwarzschild's coordinates do not extend to $r<2M$, we use the tortoise coordinate $dr/dr_*=\pm F$, where henceforth the upper and lower signs refer to the two different universes connected at the throat. Without loss of generality we assume $r_*(r_0)=0$, so that one domain is $r_*>0$ whereas the other domain is $r_*<0$. The surgery at the throat requires a thin shell of matter with surface density and surface pressure~\cite{VisserBook}
\begin{equation}
\sigma =-\frac{1}{2\pi r_0}\sqrt{F(r_0)}\,, \qquad p =\frac{1}{4\pi r_0} \frac{(1-M/r_0)}{\sqrt{F(r_0)}}\,,
\end{equation}
respectively. As required for traversable wormholes in general relativity, the weak energy condition is violated~\cite{Morris:1988tu,VisserBook}\footnote{The weak energy condition is not necessarily violated in modified gravity; e.g. in Einstein-dilaton Gauss-Bonnet gravity traversable wormholes satisfying all energy conditions exist~\cite{Kanti:2011jz}.} ($\sigma<0$), whereas the strong and null energy conditions are satisfied when the throat is within the light ring, $r_0<3M$. 

The four velocity of a particle with mass \mbox{$\mu_p\ll M$} and conserved energy $E$ in this spacetime reads \mbox{$u_p^\mu:=d x^\mu_p/d\tau =\left(E/F,\mp\sqrt{E^2-F},0,0\right)$}, where $\tau$ is the proper time, and the coordinate time $t_p$ is governed by 
%%%
\begin{equation}
 t_p'(r)=\mp\frac{E}{F\sqrt{E^2-F}}\,,\label{dtdr}
\end{equation}
%%%
where a prime denotes a derivative with respect to $r$.
An infalling object reaches the throat in finite time (we set $t_p(r_0)=0$) and emerges in the other universe. In the point-particle limit, Einstein equations coupled to the stress-energy tensor \mbox{$T^{\mu\nu}=\mu_p\int \frac{d\tau}{\sqrt{-g}} u_p^\mu u_p^\nu \delta(x^\mu-x^\mu_p(\tau))$} reduce to a pair of Zerilli equations, $\frac{d^2\psi_l(\omega, r)}{dr_*^2}+\left[\omega^2-V_l(r)\right]\psi_l(\omega,r)=S_l$, with~\cite{Zerilli:1971wd}
%%%
\begin{eqnarray}
 V_l &=& 2F\left[\frac{9 M^3+9 M^2 r \Lambda +3 M r^2 \Lambda ^2+r^3 \Lambda ^2
   (1+\Lambda )}{r^3 (3 M+r \Lambda )^2}\right]\,,\nn \\
 S_l &=& \frac{2 \sqrt{2} \mu_p E  (9+8 \Lambda )^{1/4}e^{i \omega  t_p}}{F  (3 M+r \Lambda )^2 \omega  t_p'(r)}\times\nn\\
 &&\left[F^2 {t_p'}\left(2 i  \Lambda  + (3
   M+r \Lambda ) \omega  {t_p'}\right)-(3 M+r \Lambda ) \omega\right]\,,
\end{eqnarray}
%%%
where $\Lambda=(l-1)(l+2)/2$ and $l\geq2$ is the index of the spherical-harmonic expansion. The source term is different in the two universes due to the presence of $t_p(r)$. 
The time-domain wave function can be recovered via $\Psi_l(t,r)=1/\sqrt{2\pi}\int d\omega e^{-i\omega t}\psi_l(\omega, r)$.

With the master equation in both universes at hand, we only miss the junction conditions for $\psi_l$ at the throat. The latter depend on the properties of the matter confined in the thin shell~\cite{Pani:2009ss}. For simplicity, here we assume that the microscopic properties of the shell are such that $\psi_l$ and $d\psi_l/dr_*$ are continuous at $r_*=0$. This assumption is not crucial and can be modified without changing our qualitative results.

Finally, the energy flux emitted in GWs reads~\cite{Zerilli:1971wd}
%%%
\begin{equation}
\frac{dE}{d\omega}=\frac{1}{32\pi}\sum_{l\geq2}\frac{(l+2)!}{(l-2)!}\omega^2|\psi_l(\omega,r\to\infty)|^2\,, \label{dEdw}
\end{equation}
%%%
and the solution $\psi_l$ can be obtained through the standard Green's function as
%%%
\begin{equation}
\psi_l(r)=\frac{\psi_+}{W}\int_{-\infty}^{r} dr_* S_l \psi_- +\frac{\psi_-}{W}\int_{r}^{\infty} dr_* S_l \psi_+\,,
\end{equation}
%%%
where $\psi_\pm$ are the solutions of the corresponding homogeneous problem with correct boundary conditions at \mbox{$r_*\to\pm \infty$}, and the Wronskian \mbox{$W=\psi_- d\psi_+/dr_*-\psi_+ d\psi_-/dr_*$} is constant by virtue of the field equations. We validated the results presented below by comparing this procedure with a direct integration of the master equation through a shooting method, obtaining the same results up to numerical accuracy.

%%%%%%%%%%%%%%%%%%%%%%%%%%%%%%%%%%%%%%%%%%%%%%%%%%%%%%%%%%%%%%%%%%%%%%%%%%%%%%
\noindent{\bf{\em IV. QNM spectrum.}}
%%%%%%%%%%%%%%%%%%%%%%%%%%%%%%%%%%%%%%%%%%%%%%%%%%%%%%%%%%%%%%%%%%%%%%%%%%%%%%
The QNMs of the wormhole are defined by the eigenvalue problem associated with the master equation above with $S_l=0$ and supplemented by regularity boundary conditions~\cite{Kokkotas:1999bd,Berti:2009kk,Konoplya:2011qq}. The latter are $\psi_l\sim e^{\pm i\omega r_*}$ at the asymptotic boundaries of both universes. Note that, because $r_*\to \pm r$ at infinity, in Schwarzschild coordinates both homogeneous equations and boundary conditions are the same. At the throat we impose continuity of $d\psi_l/dr_*$ which ---~given the symmetry of the problem and the homogeneity of the master equation~--- can be achieved only in two ways: by imposing either $d\psi_l(0)/dr_*=0$ or $\psi_l(0)=0$. Correspondingly, we find two families of QNMs, $\omega=\omega_R+i\omega_I$, that can be obtained by a straightforward direct integration supplied by a high-order asymptotic expansion of the solution~\cite{Pani:2013pma} in either of the two domains.
\begin{figure}[th]
\begin{center}
\epsfig{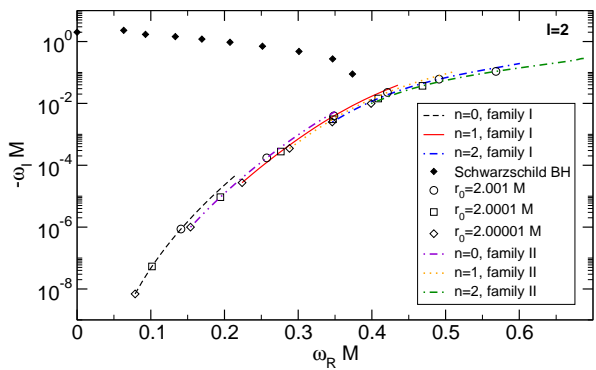}
\caption{
The first three tones $(n=0,1,2$) for the two families of polar $l=2$ QNMs of a wormhole parametrically shown in the complex plane for different values of the throat location $r_0$, and compared to the first QNMs of a Schwarzschild BH. In the BH limit ($r_0\to 2M$) all QNMs of the wormhole approach the real axis.
\label{fig:QNMs}}
\end{center}
\end{figure}

A representative example of the polar QNM spectrum is shown in Fig.~\ref{fig:QNMs}. Remarkably, in the BH limit \mbox{($r_0\to 2M$)} the spectrum is dramatically different from that of a Schwarzschild BH. While the fundamental mode of a Schwarzschild BH is $\omega_{\rm BH}M\sim0.3737 -0.0890 i$, as $r_0\to 2M$, the QNMs of the wormhole approach the real axis and become long lived, e.g. the fundamental mode is $\omega_{\rm WH}M\approx 0.0788-6.93\times 10^{-9}i$ when $r_0=2.00001M$. In fact, as $r_0\to 2M$ the deviations from the BH QNMs are \emph{arbitrarily} large. 

\begin{figure}[th]
\begin{center}
\epsfig{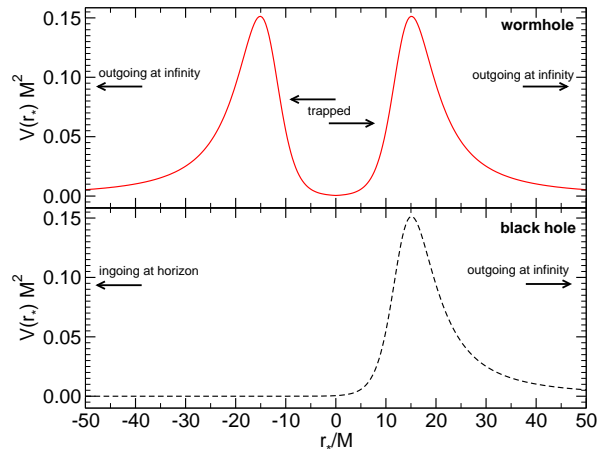}
\caption{
Effective ($l=2$) potential in tortoise coordinates for a static traversable wormhole (top panel) with $r_0=2.001M$ and for a Schwarzschild BH (bottom panel). 
\label{fig:potential}}
\end{center}
\end{figure}

This behavior can be understood by investigating the effective potential shown in Fig.~\ref{fig:potential}.
Due to the presence of the throat at $r_*=0$, the effective potential is $Z_2$ symmetric and develops another barrier at $r_*<0$. Therefore, for any $r_0\lesssim 3M$, wormholes can support long-lived modes trapped between the two potential wells near the light rings. These modes are analog to the ``slowly damped'' modes of ultracompact stars~\cite{1991RSPSA.434..449C,Chandrasekhar:1992ey,Abramowicz:1997qk} (cf. Ref.~\cite{Cardoso:2014sna} for a detailed discussion). 

% 

\begin{figure*}[ht]
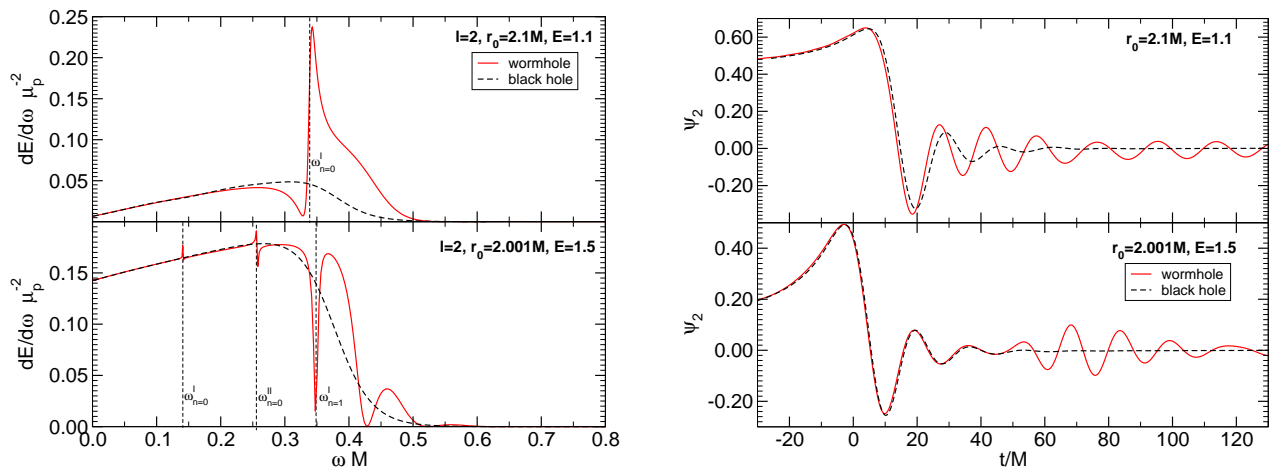

\begin{center}
\begin{tabular}{cc}
\epsfig{file=spectrum.eps,width=8.1cm,angle=0,clip=true}&
\epsfig{file=ringdown.eps,width=7.8cm,angle=0,clip=true}
\end{tabular}
\caption{
Left panels: quadrupolar GW energy spectrum [cf. Eq.~\eqref{dEdw}] for a point particle crossing a traversable wormhole and compared to the case of a particle plunging into a Schwarzschild BH with the same energy $E$. Top and bottom panels refer to $r_0=2.1M$, $E=1.1$ and to $r_0=2.001M$, $E=1.5$, respectively (different parameters give qualitatively similar results). Vertical dashed lines denote the frequency of the first QNMs of the wormhole (cf. Fig.~\ref{fig:QNMs}) which correspond to narrow resonances in the flux~\cite{Pons:2001xs,Berti:2009wx}. 
Right panels: the corresponding GW waveforms compared to the BH case. The BH waveform was shifted in time by $\Delta t$ [cf. Eq.~\eqref{Deltat}] to account for the dephasing due to the light travel time from the throat to the light ring. 
\label{fig:ringdown}}
\end{center}
\end{figure*}

%%%%%%%%%%%%%%%%%%%%%%%%%%%%%%%%%%%%%%%%%%%%%%%%%%%%%%%%%%%%%%%%%%%%%%%%%%%%%%
\noindent{\bf{\em V. Excitation of light-ring modes VS QNMs.}}
%%%%%%%%%%%%%%%%%%%%%%%%%%%%%%%%%%%%%%%%%%%%%%%%%%%%%%%%%%%%%%%%%%%%%%%%%%%%%%
Given the drastically different QNM spectrum of a wormhole relative to the BH case, one might be tempted to expect a completely different ringdown signal in actual dynamical processes.
This expectation seems to be confirmed by the energy spectrum shown in the left panel of Fig.~\ref{fig:ringdown} and compared to the case of a particle plunging into a Schwarzschild BH. The spectra coincide only at low frequencies, but are generically very different. Furthermore, in the BH limit, the long-lived QNMs of the wormhole can be excited and correspond to narrow, Breit-Wigner resonances in the spectrum~\cite{Pons:2001xs,Berti:2009wx}.

However, as previously discussed, the BH QNMs are light-ring modes and should play a role for any object with a light ring. In fact,
the striking difference in the energy spectra does not leave a trace in the initial ringdown waveform. This is shown in the right panel of Fig.~\ref{fig:ringdown} for the time-domain wave function $\Psi_2(t,r)$ extracted at infinity as a function of time. As the wormhole approaches the BH limit, $r_0\to 2M$, the initial ringdown is precisely the same as in the Schwarzschild case: the waveform
oscillates with the same fundamental QNM of a Schwarzschild BH, although the QNM spectrum of the wormhole is completely different from that of the BH. We stress that the fundamental BH QNM does \emph{not} appear as a pole of the corresponding Green's function of the wormhole, but nevertheless dominates the ringdown.

The QNMs of the wormhole contain low energy and get excited only at late times, namely after the particle crosses the throat in the characteristic time scale
%%%
\begin{equation}
 \Delta t=\int_{r_0}^{3M} \frac{d r}{F}\sim -2M \log\left(\frac{\ell}{M}\right)\,, \label{Deltat}
\end{equation}
%%%
where in the last step we considered $r_0=2M+\ell$ with $\ell\ll M$. Finally, in the BH limit ($\ell\to0$) all QNMs are long lived and have similar frequencies (cf. Fig.~\ref{fig:QNMs}), which gives rise to a peculiar beating pattern at late times.

%%%%%%%%%%%%%%%%%%%%%%%%%%%%%%%%%%%%%%%%%%%%%%%%%%%%%%%%%%%%%%%%%%%%%%%%%%%%%%
\noindent{\bf{\em VI. Discussion.}}
%%%%%%%%%%%%%%%%%%%%%%%%%%%%%%%%%%%%%%%%%%%%%%%%%%%%%%%%%%%%%%%%%%%%%%%%%%%%%%
Our results give strong evidence for a highly counterintuitive phenomenon: in the postmerger phase of a compact-binary coalescence, the initial ringdown signal chiefly depends on the properties of the light ring ---~and not on the QNMs~--- of the final object. If the latter is arbitrarily close to a BH, the ringdown modes will correspond to the BH QNMs, even if the object does \emph{not} possess a horizon. In particular, this also means that mass (and probably spin)
estimates from current ringdown templates perform well even if the compact object is horizonless.
The actual QNMs of the object are excited only at late times and typically do not contain a significant amount of energy. Therefore, they play a subdominant role in the merger waveforms, but will likely dominate over Price's power-law tails~\cite{Price:1971fb}. 

Clearly, our model is heuristic and could be extended in several ways, e.g. by including rotation, finite-size and self-force effects, and more generic orbits. None of these effects are expected to change the qualitative picture discussed above\footnote{Environmental effects (such as accretion disks, magnetic fields, dark-matter distributions or a cosmological constant) are typically negligible~\cite{Barausse:2014tra} and should not affect the waveform significantly.}.
In particular, the motion of the particle before crossing the innermost-stable circular orbit is irrelevant for the ringdown signal, which depends almost entirely on the subsequent plunge and on the particle's motion after crossing the light ring. It would be interesting to extend our analysis by performing a numerical simulation of a compact-binary merger producing a horizonless compact object.

Our results are relevant to test possible consequences of quantum effects at the horizon scale~\cite{Giddings:2016tla}, e.g. the firewall~\cite{Almheiri:2012rt} and the gravastar~\cite{Mazur:2004fk} proposals. In these models, the QNM spectrum might considerably differ from the Kerr case~\cite{Barausse:2014tra,Cardoso:2014sna}, but this will not prevent GW observatories from detecting their ringdown signal using standard BH-based templates. 
For various BH mimickers, the horizon is removed by a quantum phase transition, which would naturally occur on Planckian length scales~\cite{Mazur:2004fk,Damour:2007ap,Almheiri:2012rt,Barausse:2014tra,Giddings:2016tla}. In this case the changes to the QNM spectrum are more dramatic and, if detected, they will provide a smoking gun for quantum corrections at the horizon scale. In the $\ell\ll M$ limit, we expect that our results will be qualitatively valid for {\it any} model.
Interestingly, Eq.~\eqref{Deltat} shows that the delay $\Delta t$ for the QNMs to appear after the main burst of radiation produced at the light ring depends only logarithmically on $\ell$. For a final object with $M\approx 60 M_\odot$, $\Delta t\sim 16 \tau_{\rm BH}$ ($\tau_{\rm BH}\approx 3\,{\rm ms}$ being the fundamental damping time of a Schwarzschild BH with the same mass) even if the length scale is Planckian, $\ell\sim L_p= 2\times 10^{-33}\,{\rm cm}$. For $\ell\sim\sqrt{2 L_p M}\sim 10^{-13}\,{\rm cm}$ as in the original gravastar model~\cite{Mazur:2004fk}, such delay is only halved.

Our results suggest that future GW detections by aLIGO~\cite{aLIGO}, aVIRGO~\cite{aVIRGO}, and KAGRA~\cite{KAGRA} should focus on extracting the late-time ringdown signal, where the actual QNMs of the final object are eventually excited. Even in the absence of a horizon, these modes are expected to be in the same frequency range of the BH QNMs and therefore might be detectable with advanced GW interferometers. Furthermore, their extremely long damping time (cf. Fig.~\ref{fig:QNMs}) might be used to enhance the signal through long-time integrations, even if the energy contained in these mode is weak.
Estimating the signal-to-noise ratio required for such precise measurements is an important extension of our work.

Horizonless compact objects require exotic matter configurations and almost inevitably possess a \emph{stable} light ring at $r<3M$~\cite{Cardoso:2014sna}. The latter might be associated with various instabilities, including fragmentation and collapse~\cite{Cardoso:2014sna} and the ergoregion instability~\cite{1978CMaPh..63..243F,Cardoso:2007az,Cardoso:2008kj,Pani:2010jz} when the object rotates sufficiently fast. While our results are generic, the viability of a BH mimicker depends on the specific model, especially on its compactness and spin~\cite{Chirenti:2008pf}.

The recent GW detection by aLIGO~\cite{Abbott:2016blz} enormously strengthens the evidence for stellar-mass BHs, whose existence is already supported by various indirect observations in the electromagnetic band (cf. e.g. Refs.~\cite{Narayan:2005ie,Broderick:2005xa}). While BHs remain the most convincing Occam's razor hypothesis, it is important to bear in mind the elusive nature of an event horizon and the challenges associated with its direct detection.

The postmerger signal detected by aLIGO has been recently investigated in the context of tests of gravity (cf. e.g. Refs.~\cite{Konoplya:2016pmh,Chirenti:2016hzd}). Our results show that only late-time ringdown detections might be used to rule out exotic alternatives to BHs and to test quantum effects at the horizon scale.
As it stands, the single event GW150914~\cite{Abbott:2016blz} does not provide the final evidence for horizons, but strongly supports the existence of light rings, itself a genuinely general-relativistic effect.

% %%%%%%%%%%%%%%%%%%%%%%%%%%%%%%%%%%%%%%%%%%%%%%%%%%%%%%%%%%%%%%%%%%%%%%%%%%%%%%
\noindent{\bf{\em Acknowledgments.}}
% %%%%%%%%%%%%%%%%%%%%%%%%%%%%%%%%%%%%%%%%%%%%%%%%%%%%%%%%%%%%%%%%%%%%%%%%%%%%%%
% \begin{acknowledgments}
We thank Emanuele Berti, Valeria Ferrari and Leonardo Gualtieri for interesting comments on a draft of this Letter.
V.C. acknowledges financial support provided under the European Union's H2020 ERC Consolidator Grant ``Matter and strong-field gravity: New frontiers in Einstein's theory'' grant agreement no. MaGRaTh--646597.
Research at Perimeter Institute is supported by the Government of Canada through Industry Canada and by the Province of Ontario through the Ministry of Economic Development $\&$
Innovation.
This work was partially supported by FCT-Portugal through the project IF/00293/2013, by the H2020-MSCA-RISE-2015 Grant No. StronGrHEP-690904, and by ``NewCompstar'' (COST action MP1304).
%
% \end{acknowledgments}
%%%%%%%%%%%%%%%%%%%%%%%%%%%%%%%%%%%%%%%%%%%%%%%%%%%%%%%%%%%%%%%%%%%%%%%%%%%%%%
%\vskip 5mm

\bibliographystyle{h-physrev4}
\bibliography{Ref}

\end{document}